\documentclass[conference]{IEEEtran}

\makeatletter
\newcommand{\linebreakand}{%
  \end{@IEEEauthorhalign}
  \hfill\mbox{}\par
  \mbox{}\hfill\begin{@IEEEauthorhalign}
}
\makeatother
\ifCLASSINFOpdf
\else
\fi

\usepackage{graphicx}
\usepackage{ifthen}
\usepackage{xcolor}

\usepackage{microtype}
\usepackage{booktabs}
\usepackage{url}
\usepackage{breakurl}
\usepackage[hidelinks]{hyperref}
\def\submission{}
\def\submissionB{}

\usepackage{xspace}
\usepackage{glossaries}
\usepackage[inline]{enumitem}
\usepackage{cleveref}

\usepackage{array}
\newcolumntype{L}[1]{>{\raggedright\arraybackslash}p{#1}}

\usepackage{relsize}

\newcommand{\designnametext}{TrustMee\xspace}
\newcommand{\designname}{{\relscale{0.9}\textsf{TrustMee}}\xspace}
\newcommand{\attester}{Attester\xspace}
\newcommand{\verifier}{Verifier\xspace}
\newcommand{\verifiers}{Verifiers\xspace}
\newcommand{\component}{Verification Component\xspace}
\newcommand{\components}{Verification Components\xspace}

\newcommand{\goal}[2]{
    \vspace{0.5em}
    \par\noindent
    \newdimen\protocolstepwidth
    \protocolstepwidth=\linewidth
    \addtolength\protocolstepwidth{-2\fboxsep}
    \fbox{\begin{minipage}[c]{\protocolstepwidth}\textbf{#1}: \emph{#2}\end{minipage}}
    \par\vspace{5pt}
}

\newcommand{\change}[1]{
  \ifdefined\submission
    #1
  \else
    \textcolor{blue}{#1}
  \fi
}

\newcommand{\changeB}[1]{
  \ifdefined\submissionB
    #1
  \else
    \textcolor{blue}{#1}
  \fi
}

\newacronym{cc}{CC}{Confidential Computing}
\newacronym{csp}{CSP}{Cloud Service Provider}
\newacronym{ear}{EAR}{EAT Attestation Result}
\newacronym{eat}{EAT}{Entity Attestation Token}
\newacronym{cmw}{CMW}{Conceptual Messages Wrapper}
\newacronym{tcb}{TCB}{trusted computing base}
\newacronym{tee}{TEE}{Trusted Execution Environment}

\newacronym{as}{AS}{Attestation Service}
\newacronym{pki}{PKI}{Public Key Infrastructure}
\newacronym{rats}{RATS}{Remote ATtestation procedureS}
\newacronym{rss}{RSS}{Resident Set Size}
\newacronym{rvps}{RVPS}{Reference Value Provider Service}
\newacronym{kbs}{KBS}{Key Broker Service}
\newacronym{sev-snp}{SNP}{Secure Nested Paging}
\newacronym{tdx}{TDX}{Trust Domain Extensions}
\newacronym{td}{TD}{Trust Domain}
\newacronym{wit}{WIT}{Wasm Interface Type}
\newacronym{vtpm}{vTPM}{virtual Trusted Platform Module}
\newacronym{cvm}{CVM}{Confidential Virtual Machine}
\newacronym{epid}{EPID}{Enhanced Privacy ID}
\newacronym{psa}{PSA}{Platform Security Architecture}
\newacronym{sgx}{SGX}{Software Guard Extensions}
\newacronym{cca}{CCA}{Confidential Compute Architecture}
\newacronym{wasi}{WASI}{WebAssembly System Interface}
\newacronym{kds}{KDS}{Key Distribution Service}
\newacronym{pck}{PCK}{Provisioning Certification Key}
\newacronym{pcs}{PCS}{Provisioning Certification Service}
\newacronym{coco}{CoCo}{Confidential Containers}
\newacronym{vm}{VM}{Virtual Machine}

\begin{document}
\title{\designnametext: Self-Verifying Remote Attestation Evidence}
\ifdefined\anonymous
\author{\IEEEauthorblockN{Anonymous Author(s)}}
\else
\author{
\IEEEauthorblockN{Parsa Sadri Sinaki}
\IEEEauthorblockA{Aalto University\\
parsa.sadrisinaki@aalto.fi}
\and
\IEEEauthorblockN{Zainab Ahmad}
\IEEEauthorblockA{Ericsson Research\\
Aalto University\\
zainabahmad1511@gmail.com}
\and
\IEEEauthorblockN{Wentao Xie}
\IEEEauthorblockA{Ericsson Research\\
Aalto University\\
wentao.xie22@gmail.com}

\linebreakand

\IEEEauthorblockN{Merlijn Sebrechts}
\IEEEauthorblockA{Ghent University\\
imec\\
merlijn.sebrechts@ugent.be}
\and
\IEEEauthorblockN{Jimmy Kj{\"a}llman}
\IEEEauthorblockA{Ericsson Research\\
jimmy.kjallman@ericsson.com}
\and
\IEEEauthorblockN{Lachlan J.~Gunn}
\IEEEauthorblockA{Aalto University\\
lachlan@gunn.ee}
}
\fi
\maketitle
\begin{abstract}
  \changeB{
  Remote attestation allows a Trusted Execution Environment to attest its own state to a remote party, and is used in the Confidential Computing ecosystem to provide assurance that data and code are available only to authorized hardware and software.  This functionality can then be incorporated into higher-level platforms, allowing individual workloads to attest themselves to a remote party.
  However, supporting each new platform enlarges a Verifier's trusted computing base, since it must include code to verify the attestation evidence from each platform, often performing risky operations such as binary parsing.  This means that emerging or niche Attester platforms cannot be supported by standard Verifier software, preventing them from interacting with the broader confidential computing ecosystem.
  }
  
  We introduce the concept of \emph{self-verifying} remote attestation evidence. Each attestation bundle \change{identifies its verification logic in the form of a WebAssembly component that is downloaded by the Verifier and executed}. This approach transforms evidence verification into a \change{platform-agnostic functionality that is implemented once for all platforms}:
  \changeB{the Verifier executes the supplied verification logic to validate the evidence and incorporates its hash and, when available, the signer's public key as identity claims into the attestation result.
  }
  As a result, Verifiers can validate attestation evidence without any platform‑specific \change{code; the verification logic is just another measurement whose reference value can be checked with existing mechanisms.}
  
  We implement this concept as \designname, a platform-agnostic verification driver for the Trustee framework. We demonstrate its functionality with self-verifying evidence for AMD SNP, Intel TDX\changeB{, Intel SGX, Kata agent-policy appraisal, and ReCFA control-flow} attestations, producing attestation claims in the standard \change{Entity Attestation Token (EAT)} format. 
  \changeB{The last two are part of the software stack, and not supported by existing Verifiers such as Trustee.}
\end{abstract}
\IEEEpeerreviewmaketitle

\section{Introduction}

Remote attestation allows \verifiers to assess the integrity and authenticity of a remote device. It is a key component of \gls{cc}, allowing sensitive workloads to run within untrusted infrastructure. It achieves this by providing cryptographically-verifiable evidence that the computation runs inside a hardware-isolated \gls{tee}~\cite{coker2011principles}, and so is isolated from other workloads as well as the hypervisor and the \gls{csp} operational staff.

\begin{figure}[t]
\centering
  \includegraphics[width=\linewidth]{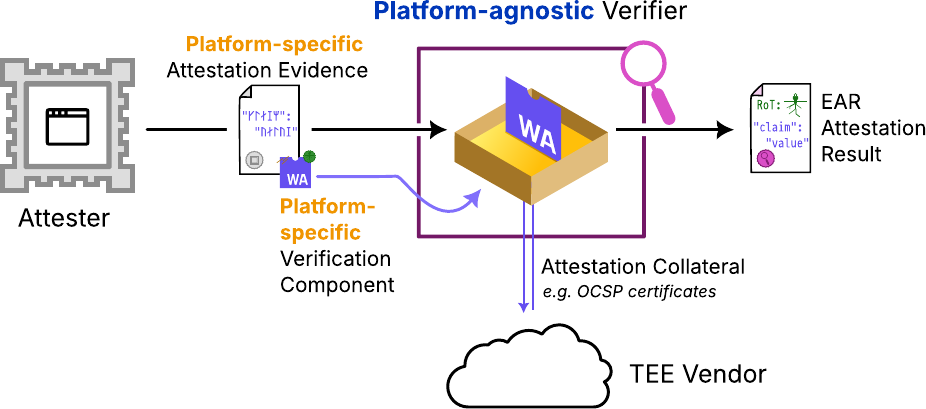}
    \caption{\designname removes the need for platform-specific code in the verifier by bundling verification logic with the evidence. A generic verifier then uses the component to process the evidence, allowing new \glspl{tee} to be supported by the verifier by just applying a new appraisal policy without any code changes.}
    \label{fig:multiplatform-attestation-architecture}
\end{figure}
\changeB{ 
That evidence is platform-specific.
Each \gls{tee} platform defines its own evidence format, protocols, and verification libraries, so a \verifier needs \gls{tee}-dependent verification logic for every \gls{tee} platform it may encounter~\cite{menetrey2022attestation}.   Additionally, attestation is not confined to the hardware layer. Deployment and software layers produce evidence of their own with their own verification procedure; e.g., a container platform might provide a statement of which container initiated an attestation, and a control-flow attestation scheme might certify a control-flow trace emitted by an instrumented binary.  These attester platforms require custom verification logic to validate this evidence and its binding to attestation evidence from an underlying \gls{tee}.
}

\changeB{
Adding support for a new Attester platform is not a one-time integration cost, and the burden does not scale merely with the number of Attester platform vendors.
Cloud deployments wrap hardware evidence in deployment-specific envelopes; for example, Trustee~\cite{confidentialcontainers2025trustee} already supports bare-metal \gls{tdx}, bare-metal \gls{sev-snp}, Azure \gls{tdx}-over-\gls{vtpm}, and Azure \gls{sev-snp}-over-\gls{vtpm} as four distinct platforms~\cite{microsoft2025azurecvmattestation}, while support for new architectures, such as CoVE for RISC-V confidential computing~\cite{sahita2023cove}, continues to add new verification logic, each of which adds to the \verifier's \gls{tcb}.
}

\changeB{
Existing frameworks, such as Veraison~\cite{ccc2023veraison} and Trustee~\cite{confidentialcontainers2025trustee}, break verification into per-platform plugins, but these plugins are native code running directly on the \verifier. This has two drawbacks: \begin{enumerate*} \item  \verifier operators must keep these plugins up-to-date for every supported Attester platform, and \item bugs or backdoors in a plugin---whose function is to parse attacker-controlled input---can compromise the \verifier as a whole. These plugins link large memory-unsafe C and C++ libraries such as OpenSSL, whose certificate parsing has historically yielded remote code execution vulnerabilities~\cite[CVE-2012-2110]{openssl-vulnerabilities}.\end{enumerate*}
A bug in any one plugin thus compromises the integrity of verification for every platform, so \verifiers cannot support every niche or emerging Attester platform.  This forces Attester platform vendors to develop their own nonstandard \verifier software, preventing them from interacting with existing infrastructure that is not adapted to their platform.
}

We propose \designname, a \emph{platform-agnostic} attestation verifier, shown in Figure~\ref{fig:multiplatform-attestation-architecture}, that \emph{eliminates the need for platform-specific verification logic} without sacrificing security or performance.

\designname takes advantage of WebAssembly to enable \emph{self-verifying} attestation evidence. Instead of hardcoding the \verifier with hardware-specific verification logic for each Attester platform, \changeB{the Attester identifies} a \change{WebAssembly component} that implements its platform-specific verification logic, and \designname then combines the component's verification output with platform-agnostic processing to return a signed attestation result in the standard \gls{ear} format~\cite{ietf-rats-ear-03}.
The \verifier service then only needs to \begin{enumerate*} \item \change{measure the component provided by the attestation request} and \item execute it inside the WebAssembly sandbox\end{enumerate*}.
All platform-specific parsing and cryptographic checks are delegated to this verification component, which is sandboxed by the runtime, \change{while the \verifier treats the component itself as another measured artifact whose signer and hash claims can be checked against reference values,} thus eliminating the ability of a compromised \verifier plugin to impersonate other platforms, or to compromise the \verifier as a whole.

\changeB{
The \designname host contains no platform-specific code, so the marginal \gls{tcb} cost of supporting a new platform is zero, and supporting one means publishing a signed component rather than upgrading every \verifier. 
We demonstrate this with components for AMD \gls{sev-snp}~\cite{christerattestation, powell2022amdsnpatttestation}, Intel \gls{tdx}~\cite{intel2022tdx}, and Intel \gls{sgx}~\cite{costan2016intel}.
In addition to hardware platforms, \designname also allows other parties to provide verification logic for software-level attestation schemes.
We demonstrate this with two additional components for schemes not directly supported by existing \verifier frameworks.
The first is a layered component for \gls{coco}~\cite{confidentialcontainers} that verifies the underlying \gls{tdx} or \gls{sev-snp} evidence and then appraises the Kata~\cite{randazzo2019kata} agent policy bound to the \gls{cvm}, exposing the permitted container images, execution commands, and debug settings as claims in the attestation result, where Trustee today exposes only the opaque policy digest. The second is a port of the ReCFA~\cite{zhang2021recfa} control-flow attestation \verifier, which replays an execution trace against an expected control-flow graph, requiring custom scheme-specific verification logic in existing \verifiers. Neither requires any change to the \designname \verifier. 
In our evaluation, warm-start latency matches Trustee's shipped \gls{tdx} and \gls{sgx} paths, and \gls{sev-snp} overhead falls below 3 ms with host-accelerated cryptography.
}

Our contributions are as follows:
\begin{itemize}
  \item \designname: an architecture for self-verifying remote attestation evidence, in which the attesting \changeB{environment} supplies the code needed to verify its attestation evidence. \changeB{A \verifier can therefore support a new platform, or a new attested layer, without hardcoding platform-specific logic. This eliminates the incentive for \verifiers to minimize platform support for the sake of security and deployability}.
  \item An implementation of \designname for the Trustee framework, along with verification components for AMD \gls{sev-snp}~\cite{christerattestation, powell2022amdsnpatttestation}, Intel \gls{tdx}~\cite{intel2022tdx}, and Intel \gls{sgx}~\cite{costan2016intel} attestation evidence\changeB{, as well as a layered \component that semantically appraises Kata Containers agent policies, and a \component that adopts the ReCFA~\cite{zhang2021recfa} control-flow attestation \verifier}.
  \item Security and performance analysis of \designname, demonstrating that \designname can securely verify attestation evidence with low overhead.
\end{itemize}

\section{Background}

\subsection{Remote Attestation}
\label{sec:ra}

Remote attestation is a security mechanism in which a prover (Attester) sends evidence of its state to a remote verifier~\cite{coker2011principles}. In trusted computing systems, the prover collects measurements of its configuration and reports them as evidence so the verifier can assess authenticity and integrity~\cite{bartock2015trusted}. In \gls{cc} and \glspl{tee}, this is often essential to ensure the communication endpoint is a genuine node running a trusted configuration, such as specific software inside a \gls{tee}~\cite{alder2019s}. When the prover is software running in a \gls{tee}, it typically derives cryptographic measurements of the code and other \gls{tee}-specific digests, then signs these claims with a hardware-protected key to produce evidence for verification. After successful verification, a client can safely provision secrets or sensitive workloads to the \gls{tee}, without needing to trust the underlying host environment to avoid unauthorized access or tampering~\cite{knauth2018integrating}.

\begin{figure}[t]
\centering
  \includegraphics[width=\linewidth]{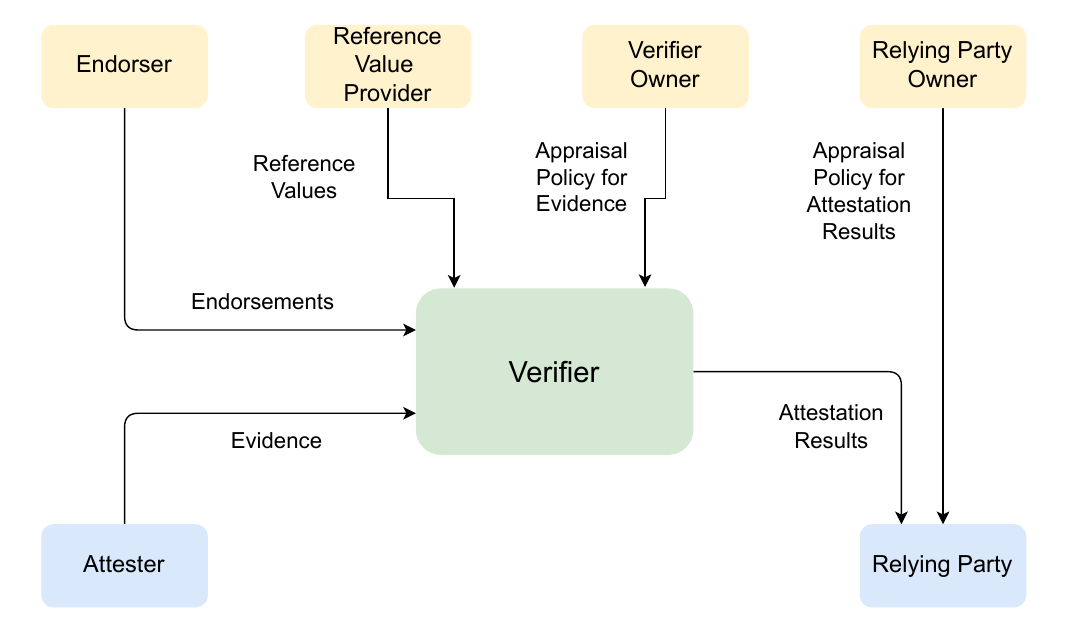}
  \caption{ The \gls{rats} architecture data flow (adapted from~\cite{birkholz2023remote}).}
  \label{fig:rats-architecture}
\end{figure}

\gls{rats} is an IETF framework that has been developed to provide a general model and terminology for remote attestation~\cite{birkholz2023remote}.
We follow the IETF \gls{rats} model for remote attestation. An Attester running inside a \gls{tee} produces signed evidence about its configuration; a Verifier appraises this evidence using endorsements and reference values; and a Relying Party consumes the resulting appraisal. Additional roles such as Endorsers and Reference Value Providers supply trusted input to the Verifier. The Verifier Owner is the entity that configures the appraisal policy for the Verifier and the Relying Party Owner is the entity that manages the appraisal policy for the Relying Party.
Figure~\ref{fig:rats-architecture} summarizes this data flow.

\textbf{\change{Entity Attestation Tokens.}}
\glsreset{ear}
\change{The \gls{eat} from RFC 9711~\cite{rfc9711} is a standardized format for encoding claims about an entity, suitable for representing attestation evidence.}
The \gls{ear}~\cite{ietf-rats-ear-03} is \change{an \gls{eat} token}, augmented with the \verifier's appraisal of an \attester's evidence. It embeds a normalized trustworthiness vector and contextual metadata bound to the appraisal process, making it easier for Relying Parties to interpret and enforce policies reliably and consistently.

\textbf{\change{Conceptual Messages Wrapper.}}
\change{The \gls{cmw} from RFC 9999~\cite{rfc9999} defines a common format for carrying different data units between \gls{rats} roles.  It allows evidence to be encapsulated alongside endorsements and reference values to be used during verification.}

\subsection{Trustee}

We build on Trustee~\cite{confidentialcontainers2025trustee}, an open-source attestation framework\change{, part of the Confidential Containers project}. Trustee includes
an Attestation Service, and a \gls{rvps}. The Attestation Service implements the \gls{rats} Verifier role.
In this work, we extend the Attestation Service component.
The components in Trustee and the interaction between them follow the definitions in the \gls{rats} architecture.

\subsection{WebAssembly}

WebAssembly is a compact, portable bytecode format that runs in a sandboxed execution environment with near-native performance~\cite{haas2017bringing, jangda2019not}. A single WebAssembly binary can execute across different operating systems and architectures as long as a compatible runtime is available, which makes it attractive as a carrier for platform-specific verification logic~\cite{haas2017bringing, webassembly2025spec}. Runtimes expose standard system interfaces through \gls{wasi}, allowing controlled access to I/O (e.g., HTTP, filesystem) while preserving isolation~\cite{webassembly2025wasi,  wasi2023website}.
The WebAssembly Component Model further structures applications as components that interact via typed interfaces defined in the \gls{wit} language~\cite{bytecodealliance2025witreference, bytecodealliance2024componentintro}. \gls{wit} lets us describe rich data types and function signatures in a language-agnostic way, so components written in different languages can exchange structured data without ad-hoc serialization. 

Prior work shows that WebAssembly provides strong isolation and near-native performance~\cite{lehmann2020everything, herrera2018webassembly, rossberg2016webassembly, konoth2018minesweeper, zhang2025research}; we leverage these properties to safely execute untrusted platform-specific verifiers.

\section{Problem description}

\subsection{System model}

\designname functions as a \gls{rats} \emph{Verifier}, accepting  attestation evidence produced by an \emph{Attester}---a node running in a \gls{tee}---verifying its validity, checking its claims against some predefined policy, and outputting the result for use by a \emph{Relying Party}.
A \gls{tee} node runs a workload (the code being attested) inside a hardware-isolated environment (e.g., a \gls{cvm}).

\changeB{
\textbf{Ownership.}
Each role owns a distinct part of the system.
The \emph{Attester platform vendor} (e.g., a \gls{tee} vendor) develops and signs the per-platform \component; this mirrors existing practice, where platform-specific attestation logic is packaged in dedicated libraries, such as Intel's DCAP libraries~\cite{intel2019sgxtdxdcapquotevl}, or provided by trusted non-vendor projects, such as VirTEE's SEV tooling~\cite{virtee_sev}. 
Instead of distributing this logic for integration into each \verifier, \designname allows the developer to distribute the \component to Attesters, which bundle or reference it with their evidence; it does not author verification logic.
The \emph{\verifier operator} owns the platform-agnostic host and the WebAssembly runtime. The host measures the \component, the runtime sandboxes and executes it, and the \verifier applies the appraisal policy, and signs the \gls{ear} result. The operator therefore owns the host \gls{tcb} and the result-signing key.
The \emph{Relying Party} receives the signed result, including the \component-identity claim, and makes the final trust decision against its policy.
}

\subsection{Adversary model}
\label{sec:adv-model}
We consider adversaries who attempt to make \designname accept invalid attestation evidence, or who attempt to compromise \designname by using malicious or vulnerable verification logic.

\textbf{Capabilities.}
The adversaries may control the network and run arbitrary software in the Attester's workload environment. 
The cloud provider may control the \changeB{Attester's} host OS, hypervisor/VMM, and networking, but cannot manipulate the attestation functionality in such a way as to obtain attestations that do not accurately represent the state of the Attester.
\changeB{The Verifier Host is not controlled by the adversary.}

\textbf{Assumptions.}
We assume the \gls{tee}'s attestation mechanism is sound, meaning that the \gls{tee} cannot be manipulated to produce valid attestations that make false claims, such as by signing false measurements or \gls{tcb} information.
We assume that authorized attestation verification logic is semantically correct for its platform, meaning that it correctly validates evidence and extracts the corresponding claims. We do not assume that this logic is free of implementation vulnerabilities that could otherwise compromise a native Verifier.
We assume also that standard cryptography is secure, as is the public key infrastructure used to certify hardware roots of trust and any code provided by the Attester.

We assume that an adversary cannot compromise the \verifier or Relying Party, nor can they escape the isolation of the WebAssembly runtime.
\changeB{We evaluate the WebAssembly runtime and compiler as trusted components in Section~\ref{sec:security-evaluation}.}

\changeB{
\textbf{Trust boundaries.}
The host-security \gls{tcb} comprises the platform-agnostic Verifier Host, including policy enforcement and result signing, and the WebAssembly runtime and compiler. The \components are outside this \gls{tcb} from the perspective of host compromise. However, an authorized \component remains part of the result-correctness \gls{tcb} for the platform whose evidence it validates.
}

\textbf{Untrusted inputs.}
We treat \begin{enumerate*} \item the received attestation evidence, \item \changeB{endorsements}, and \item the WebAssembly \component \end{enumerate*} as attacker-controlled.

\subsection{Requirements}
\label{sec:requirements}
The goal of this research is to ensure that a \verifier can validate attestation evidence without any platform-specific code, while still supporting the current diversity in verification protocols.
Therefore, the following requirements must be fulfilled by \designname:
    \goal{\hypertarget{req.CR}{CR}---Compatibility}{
    Any Verifier must be able to verify the attestations from any Attester platform, including layered attestation schemes, without any Attester-specific modifications to the Verifier, or any Verifier-specific modifications to the Attester.
    }
    Attester-specific unsandboxed native Verifier drivers create a large and rapidly changing verifier-side \gls{tcb}, and require Verifier updates before new platforms or schemes can be used.
    A platform-agnostic \verifier instead reduces trusted code and enables instant deployment of new Attester platforms without modifying or redeploying the Verifier.
    \changeB{Trustee~\cite{confidentialcontainers2025trustee} and Veraison~\cite{ccc2023veraison} do not meet this requirement because both require \verifier-side modifications to support new Attester platforms.}
    \goal{\hypertarget{req.SR}{SR}---Security}{
    Attestation results as determined by the Verifier must reflect the true state of the Attester.
    }

    Verifying attestation evidence involves a number of tasks such as binary parsing of certificates that are historically error-prone, and thus platform-specific \verifiers may have exploitable vulnerabilities.  This is exacerbated by the fact that each \gls{tee} vendor must produce their own libraries supporting each of their platforms, meaning that even open-source verifier libraries will not be subject to the sustained and concentrated scrutiny that a platform-agnostic library may see.
    \changeB{To contain the impact of such vulnerabilities, adding support for a new platform in the \verifier must not enlarge the \verifier \gls{tcb} or the \gls{tcb} of existing platform-specific verifiers.}
    \goal{\hypertarget{req.PR}{PR}---Performance}{
    When implemented in existing attestation frameworks, the design should add minimal end-to-end latency compared with existing drivers.
    }
    Attestation is often on the critical path of connection establishment or workload admission (e.g., TLS handshakes, gateway decisions, orchestrator scheduling), therefore making
    low latency a key requirement for real-world adoption.

\section{Self-Verifying Remote Attestation Evidence}
\label{sec:self-verifying-evidence}

\designname serves as a \gls{rats} Verifier.
\designname receives evidence from the Attester, interacts with the Endorser, Reference Value Provider and Verifier Owner to obtain additional data needed to verify the validity of the evidence and the appropriateness of the Attester's state, then outputs the attestation result to the Relying Party.

\begin{figure*}[h]
    \centering
    \includegraphics[width=\linewidth]{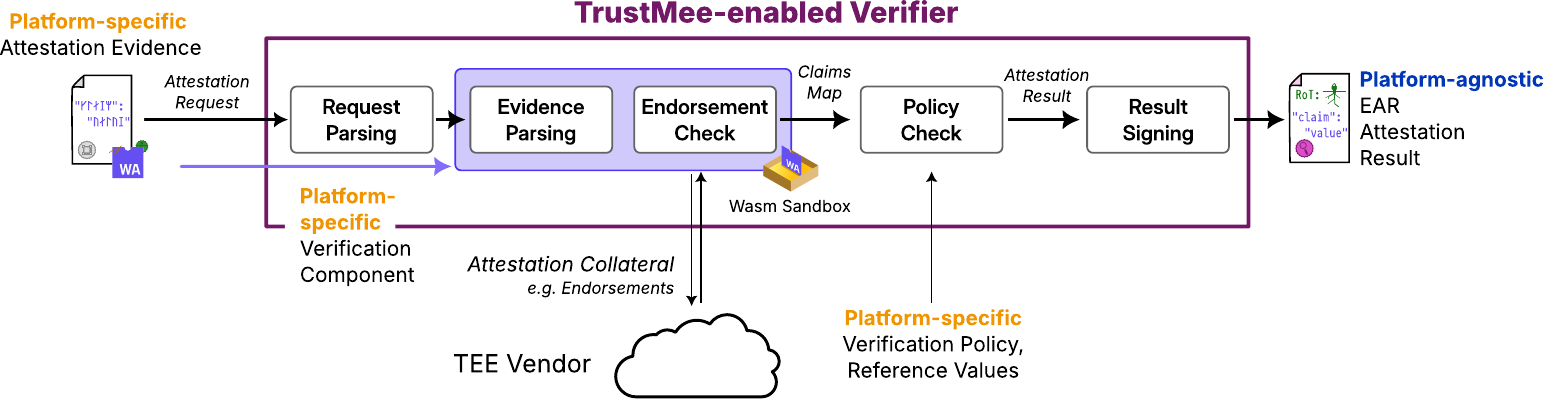}
    \caption{\designname architecture for self-verifying remote attestation evidence. The platform-specific verification logic is bundled with the evidence. \designname executes it as a WebAssembly component inside a sandboxed WebAssembly runtime to map the platform-specific attestation evidence to \change{claims} that can be used by the platform-agnostic logic of the \verifier.}
    \label{fig:attestation-architecture}
\end{figure*}

Figure~\ref{fig:attestation-architecture} illustrates the design of \designname.  Attestation verification involves the following steps:
\begin{itemize}
  \item Request Parsing (platform-agnostic code, platform-specific data): Parse the attestation request to obtain the evidence collected from the Attester, the claims expected to be found in the attestation, a policy ID to specify which policy to check against, and the \component to be loaded.
  \item Evidence Parsing (platform-specific code, platform-specific data): Parse the evidence, which varies in format across different Attester platforms.
  \item Endorsement Check (platform-specific code, platform-specific data): Check the evidence against the endorsement, including signature and certificate validation where applicable. For hardware-backed evidence, these endorsements are typically obtained from the \gls{tee} manufacturer.
  \item Policy Check (platform-agnostic code, platform-specific data): Fetch the reference values from the \gls{rvps} and check the parsed claims against them according to the assigned policy from the request.
  After parsing and validating the evidence, the extracted values must be appraised against trusted reference values under policy; this step is partly platform-specific because the claim set and its semantics are derived from platform-specific evidence formats, with some fields following vendor-defined conventions (e.g., Intel/AMD measurement and TCB-related fields) and others being verifier-defined.
  \item Result Signing (platform-agnostic code, platform-agnostic data): Sign the attestation result with the signing key of the \verifier.
\end{itemize}

\designname extracts the steps with platform-specific code into a WebAssembly \emph{\component} that can be executed by a platform-agnostic \designname-enabled Verifier. The \component exposes an attestation API that remains consistent across all Attester platforms.

Self-verifying attestation evidence consists of two parts: the Attester's platform-specific attestation evidence, and the \component.  \change{The \verifier is responsible only for running the component in the WebAssembly sandbox, then checking the result against a policy, and signing the result.} This makes multi-platform attestation verification \change{a platform-agnostic functionality that is implemented and deployed once for all platforms}, avoiding the need to update the Verifier as new Attester platforms emerge.

This design offers several advantages \changeB{over state-of-the-art \verifiers such as Trustee~\cite{confidentialcontainers2025trustee}}:
\begin{enumerate}
  
  \item \textbf{Rapid support for new \glspl{tee} and updates.}
  \changeB{The \verifier Host implements only platform-agnostic request handling, policy evaluation, and result signing, while evidence parsing and endorsement validation are delegated to WebAssembly \components. New or updated \components can therefore support new attestation evidence formats and validation rules without modifying or redeploying the \verifier, provided that they implement the required interface.
  }
  
  \item \textbf{Stronger isolation for error-prone parsing and cryptographic operations.}
  Platform-specific evidence parsing and certificate processing run inside the WebAssembly sandbox behind a constrained host interface, limiting the impact of malformed evidence and vulnerabilities in \verifier code.

  \item \changeB{\textbf{Verification of layered and composite evidence.} 
  Because every \component imports and exports the same interface, a \component for a higher layer can invoke the \component for the layer beneath it and be composed with it into a single unit. Evidence spanning several layers, such as a deployment policy bound to a \gls{cvm}'s hardware measurements, can therefore be appraised as a whole, by logic supplied by the party that owns the upper layer, rather than by claim-matching rules hardcoded in the Verifier (Section~\ref{sec:kata-verifier}).}

  \item \changeB{\textbf{Visibility of the verification logic to the Relying Party.} 
  The \verifier Host emits the \component's hash and, when available, the signer's public key as claims in the signed attestation result. A Relying Party can therefore decide whether to trust the logic that produced an appraisal, rather than implicitly trusting whichever plugin set the \verifier operator has deployed, as it must today.}
\end{enumerate}

\subsection{\component}
The main tasks of the \component are evidence parsing and endorsement checking, which together validate the evidence according to the corresponding attestation scheme. Although its internal implementation differs across platforms and schemes, the \component exposes a consistent interface, without concern for its internal implementation details.

We implement \components using the WebAssembly component model; since WebAssembly components are strongly isolated from the outside world and communicate only through well-defined interfaces, this allows \components to be loaded and executed without concern that a vulnerable \component may compromise the \verifier as a whole.

\label{sec:wit}

\begin{figure}[h]
    \centering
    \fbox{
    \includegraphics[width=0.7\linewidth]{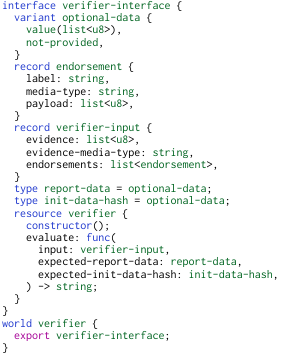}}
    \caption{The interface between the \component and \verifier Host is defined using \gls{wit} to ensure interoperability.}
    \label{fig:wit}
\end{figure}

The \component's interface is shown in \Cref{fig:wit}.  The main logic of the component is the \texttt{verifier} resource, providing a handle to a single instance of the verification logic.
In addition, a component can import \gls{wasi} interfaces to access host functionality from the runtime, such as clocks and the filesystem.  The component's \texttt{evaluate} method validates attestation evidence and returns the claims by the \attester to be checked against the policy.

\subsection{\verifier Host}
The \component is loaded and executed by the \verifier Host, which implements the platform-agnostic \verifier functionality, exposing an API that accepts attestation requests.  This API accepts a WebAssembly \component binary, attestation evidence, and a verification policy, and returns a signed response containing the attestation result in EAR format.

\change{
Before loading a WebAssembly \component, the \verifier Host checks for an optional signature. 
\changeB{It always emits the \component hash as an identity claim and, if a signature is available, also emits the signer's public key. 
Appraisal policy must authorize these identity claims. The policy and reference values used for appraisal must be selected independently of claims produced by the \component. For example, if a \component returns a claim identifying the evidence as \gls{tdx}, the \verifier must not use that claim to select the \gls{tdx} policy or reference values, since a malicious \component could emit the same claim. The selected policy must instead check that the \component identity established by the Verifier Host is permitted for that policy.
}
Authorizing a signer rather than a specific component hash allows an appraisal policy to accept any component signed by that signer---e.g.,\ a well-known \gls{tee} vendor---without updating its reference values when the component changes.}

\change{
Each trusted public key in the trust store is associated with an execution policy that bounds the \component's computation and network access. Components signed by that key run under their execution policy, whereas unsigned components and components signed by unknown keys are still accepted, but run under a restrictive default policy with a bounded computation budget and no network access. These restrictions protect the \verifier Host against denial of service, network abuse, access to other platforms' state, disclosure of its result-signing key, and the use of the \verifier itself as a launching point for further attacks.
}

\change{
Network access is unnecessary because endorsements, such as collateral, can be stapled to the attestation verification request, including them alongside the evidence in a \gls{cmw} collection. The \verifier Host extracts these endorsements from the request's \gls{cmw} collection and passes them to the \component.
}
\changeB{
The two mechanisms serve different parties. The \component's identity allows the Relying Party to decide whether to trust the verification logic, while sandboxing protects the \verifier operator when executing that logic. A Relying Party's trust in a \component's result therefore does not remove the need to isolate it from the host and from other verification operations.
}

\section{Implementation}
\label{sec:impl}

\subsection{\designnametext driver for Trustee}

\changeB{
Trustee supports various \gls{tee} platforms through a separate \verifier Driver for each platform, with each driver verifying evidence for that specific \gls{tee}. We add \designname as a new platform-agnostic \verifier Driver without changing the Attestation Service API. Rather than embedding platform-specific verification logic in the driver, \designname delegates it to dynamically loaded \components, while Trustee’s RVPS and Policy Engine continue to fetch reference values and appraise the returned claims. We provide \components for AMD \gls{sev-snp}, Intel \gls{tdx}, Intel \gls{sgx}, layered Kata policy appraisal, and ReCFA control-flow attestation.
\footnote{The \designname implementation and evaluation artifacts are available
at \url{https://anonymous.4open.science/r/trustmee-artifact-BFB9}.}
}

We sign each \component we provide using wasmsign2~\cite{wasmsignatures2024wasmsign2}, which embeds a signature section in the component binary and allows the Verifier Host to authenticate its signer. Each signature carries an expiry time that bounds its validity window, limiting rollback attacks in which an attacker replays a previously signed \component that has since been found vulnerable. \changeB{For further protection, the \verifier may enforce revocation checking, such as OCSP~\cite{rfc6960}, to reject a revoked \component before its signature expires.}

\subsection{\changeB{\components for hardware platforms} }
As Trustee already implements Verifier Drivers for AMD \gls{sev-snp}, Intel \gls{tdx}, and Intel \gls{sgx}, we extracted the key functionality of these drivers and ported them to the WebAssembly target, exposing the interface described in \Cref{sec:wit}.
The \components reuse the Trustee Verifier Driver code to the greatest extent possible; however, in some cases changes were needed in order to replace dependencies that did not support their toolchain's WebAssembly target.
\changeB{
In these cases the non-WebAssembly-compatible dependencies are replaced or adapted.
}

\subsubsection{AMD SEV-SNP Verifier}

For AMD \gls{sev-snp}, the WebAssembly verifier parses \gls{sev-snp} evidence, validates the embedded signature against a certificate chain from AMD, checks configuration fields against expected values, and exports a claims map for policy checking.
The certificates \change{can be stapled to the attestation request, or fetched directly from AMD's \gls{kds} by a \component that has been granted network access}.

\changeB{
The Trustee driver relies on the Rust \texttt{openssl} crate, whose bindings normally target a native OpenSSL installation and therefore do not compile directly to WebAssembly.
To address this, we compile OpenSSL as a WebAssembly library and link the Rust \texttt{openssl} bindings against it, enabling signature verification inside the component (with SIMD128 enabled for speed).
}

\subsubsection{Intel TDX Verifier}
\label{sec:tdx-verifier-design}
The Intel \gls{tdx} \component parses the \gls{td} Quote from a \gls{tdx} guest, validates its signature and certificate chain using Intel DCAP collateral obtained from Intel at runtime (e.g., \gls{pck} certificate chain, revocation information, and \gls{tcb}-related metadata), checks the quote's reported measurements against expected values, and emits a claims map for policy checking.

Because Intel's DCAP quote verification library~\cite{intel2025sgxdatacenter} could not easily be ported to WebAssembly, we replaced it with the pure-Rust \texttt{dcap-qvl} library~\cite{phalanetwork2025dcapqvl} and extended it to use \texttt{waki} for collateral fetching.

\subsubsection{Intel SGX Verifier}
\label{sec:sgx-verifier-design}

The Intel \gls{sgx} \component has no new porting challenges. Similar to the \gls{tdx} \component, we replace Intel's DCAP quote verification library calls with the pure-Rust \texttt{dcap-qvl} crate to compile the verifier to WebAssembly, and use the WASI-backed \texttt{waki} library for collateral fetching.

\subsection{\changeB{\components for software platforms}}

\subsubsection{\changeB{Kata Containers Policy Verifier}}
\label{sec:kata-verifier}

\changeB{
To demonstrate higher-layer verification, we implement a layered \component for the \texttt{genpolicy}-generated~\cite{kata2025genpolicy} Kata agent policy stored in \gls{coco}~\cite{confidentialcontainers} \gls{cvm} init-data~\cite{coco2025initdata}. This policy controls which container images, execution commands, and debug facilities the untrusted host may request. 
}

\changeB{
The \component imports and exports the same \gls{wit} interface described in \Cref{sec:wit}, and is statically composed with a lower-level hardware \component, \gls{tdx} or \gls{sev-snp}, yielding a single WebAssembly component.
It verifies the hardware evidence, requires the init-data digest to match the binding register (\texttt{MRCONFIGID} on \gls{tdx}, \texttt{HOSTDATA} on \gls{sev-snp}), which binds the policy to the \gls{cvm}. 
}

\changeB{
A \texttt{genpolicy}-generated policy consists of a static Rego rule template loaded from the selected \texttt{rules.rego} file, followed by a \texttt{policy\_data} assignment containing a deployment-specific JSON object. \texttt{genpolicy} copies the rule template unchanged, so policies generated using the same \texttt{rules.rego} file have identical static bytes, while \texttt{policy\_data} varies by deployment. Kata releases normally distribute a default \texttt{rules.rego} alongside \texttt{genpolicy}, although this file may change between releases and users may select a custom rules file.
The \component hashes the static part and compares it with a list of approved rule templates supplied as a stapled endorsement. Then it parses the dynamic JSON object and emits permitted container images, allowed exec commands, and debug-related flags as claims. The workload can therefore change images or commands without registering a new policy digest per deployment, while a rule not matching an approved template is identified for final policy appraisal.
}

\changeB{
Invalid hardware evidence, binding mismatches, and malformed policy structure are hard failures.
In these cases, the \component returns a verification failure rather than claims for final policy appraisal.
In contrast, a well-formed policy whose static rules do not match an approved rule template is not rejected by the \component. The \component still emits the extracted high-level policy claims, together with the rule-template hash and an \texttt{approved\_rule\_template: false} claim, allowing the final appraisal policy to decide whether the policy is acceptable. 
Trustee cannot currently support this appraisal without modification because its Rego policy receives only claims produced by the Verifier Driver, including the opaque policy digest, rather than the policy itself.
Supporting this appraisal in Trustee would require a native Verifier Driver to separate the static rules from the deployment-specific data, check whether the rules are approved, and expose the extracted high-level claims to Rego.
}

\subsubsection{\changeB{Control-Flow Attestation Verifier}}
\label{sec:recfa-verifier}

\changeB{
ReCFA~\cite{zhang2021recfa} is a control-flow attestation scheme in which the evidence is a condensed trace of the \attester's runtime control-flow events, emitted by a statically instrumented binary. Verification replays the trace against a policy with a shadow stack, identifying any violating edge. It differs from a hardware quote verification because its algorithm is stateful, and its reference data is large and structured rather than a set of expected measurements.
}

\changeB{
We compile the ReCFA verifier to \texttt{wasm32-wasip2} with wasi-sdk. The required changes include obtaining headers indirectly, replacing filesystem reads with in-memory inputs, and converting process-terminating error paths into returned verdicts.
}

\changeB{
The disassembly of the attested binary, its control-flow graph, and the two policy mappings that describe legal indirect-call targets and filtered-out call sites, are alongside the evidence. The \component emits the verdict, event count, violating edge, and digests of all four inputs as claims, so a policy can pin the exact data used.
}

\changeB{
These four inputs are conceptually reference values.
ReCFA requires these reference data during verification. Trustee’s current Verifier–policy interface cannot supply them to a Verifier Driver, so adding a new Verifier Driver alone is insufficient. More generally, \components need access to reference values for checks that cannot be deferred to final policy appraisal. Our implementation therefore supplies them alongside the evidence.
}

\changeB{
Because ReCFA evidence carries no hardware root of trust, the \component reports this as an explicit claim, and binds the trace to a supplied expected report data value. We do not implement this composition for ReCFA, but the ReCFA \component can be composed with a \gls{tdx} or \gls{sev-snp} \component to provide a hardware anchor, following the same approach used by the Kata policy \component (\Cref{sec:kata-verifier}).
}

\subsection{\verifier Host}

\change{The \designname-enabled Trustee Attestation Service exposes the Attestation API. The Attestation API is unmodified from its original implementation in Trustee.}

\subsubsection{Attestation Service API}
\label{sec:api}
\change{
The Attestation Service exposes a single \emph{Attestation API} that accepts an attestation request and outputs a signed \gls{ear} as the attestation result. Requests are wrapped in a \gls{cmw} collection. 
\changeB{\designname encodes its evidence payload using CBOR, whereas Trustee's existing native request path represents binary evidence in JSON using Base64 encoding.}
A \designname \gls{cmw} carries \begin{enumerate*} \item \designname evidence, which includes the attestation evidence and an identifier of the \component expected to process it; \item optionally, one or more stapled endorsements, e.g., \gls{tdx} collateral or \gls{sev-snp} VCEK/VLEK certificates; and \item optionally, the raw bytes of the \component itself\end{enumerate*}.
}

\changeB{
The \component identifier is transport-agnostic. The Verifier Host first uses an already-compiled local cache entry. If none exists, it loads a \component stapled in the \gls{cmw}, compiles it, and places it in the cache. If no component is stapled, it instead fetches the referenced artifact from an untrusted OCI registry such as Docker Hub, then compiles and caches it. Because the component measurement is included in the output, neither the registry nor the binary need be trusted a priori, so a developer such as a \gls{tee} vendor can publish under a well-known tag. Before instantiation, the Host evaluates the component against its trust store to select the execution policy for computation and network access. Caching avoids repeated network transfer and compilation.
}

\subsubsection{Hosting the WebAssembly \component}

To invoke the platform-specific verification functionality encapsulated in the WebAssembly \component, the \component is instantiated using the Wasmtime~\cite{bytecodealliance2022wasmtime} WebAssembly runtime. \change{We configure Wasmtime with a bounded execution budget and a restricted set of \gls{wasi} interfaces as follows}:
\begin{enumerate}
    \item \change{Fuel metering: We enable Wasmtime's fuel consumption mechanism using \texttt{Config::consume\_fuel}~\cite{wasmtime-docs} to bound the amount of computation that a \component can perform. The fuel budget is selected according to the \component's execution policy. If the budget is exhausted, \component execution is aborted and the attestation request is rejected.}
    \item WASI-HTTP: The WASI-HTTP interface is enabled to allow the WebAssembly \component to fetch endorsements from external sources\change{, but only when the \component's execution policy permits network access}.
    \item WASI-Filesystem: A local directory is exposed to each \component instance, and a handle to that directory is made available to the WebAssembly component to enable the collateral caching we implemented for the \gls{tdx} and \gls{sev-snp} Verifiers; each \component's filesystem access is isolated from other instances.
\end{enumerate}

\subsection{\change{\designnametext Policy and Attester Tools}}

\change{
For policy appraisal, we provide a tool that receives an existing Trustee policy and generates a corresponding \designname policy. The generated policy adds the required reference value checks for the new claims, including the \component hash and the signer's---normally the \gls{tee} vendor's---public key. In addition, the tool rewrites the claim paths used by the remaining policy rules to match the structure of the nested \gls{eat} token returned by the \component.
}

\change{
The only change made to the Attester is the development of a library that receives raw attestation evidence and wraps it for consumption by \designname.
}

\section{Evaluation}

\subsection{Compatibility Evaluation}

The primary requirement of \designname is Compatibility (CR, Section~\ref{sec:requirements}), meaning a \verifier must be able to validate attestations from different Attester platforms without integrating platform-specific verification code.
\changeB{
The same unmodified \designname Host dynamically loads \components implementing the common \gls{wit} interface and verifies \gls{sev-snp}, \gls{tdx}, \gls{sgx} hardware evidence, as well as Kata policy, and ReCFA control-flow evidence end to end. 
The common interface also makes layered and composite verification straightforward to support by composing \components rather than modifying the \verifier Host.
Supporting additional platforms therefore requires only distribution of an optionally signed \component, not modifying or redeploying the \verifier Host, satisfying (CR, Section~\ref{sec:requirements}).
}

\subsection{Security Evaluation}
\label{sec:security-evaluation}

\change{
The second requirement of \designname is Security (SR, Section~\ref{sec:requirements}). We evaluate \designname against the adversary model in Section~\ref{sec:adv-model}.
}

\changeB{\textbf{Evidence and endorsements.}}
\changeB{
Modified evidence fails the \component's signature and certificate validation; replayed, replaced, or misbound evidence fails freshness or report-data binding checks. Endorsements are untrusted and are accepted only after validating their signatures, chains, and correspondence to the evidence. Acceptance of a false state therefore requires failure of the underlying \gls{tee}, cryptography, \gls{pki}, freshness mechanism, or appraisal policy.
}

\changeB{\textbf{\components.}}
\change{
A malicious or vulnerable \component is also treated as attacker-controlled input. 
If the adversary modifies or replaces the \component, this does not by itself make the attestation result acceptable because the \verifier Host emits the \component hash and, when available, the signer's public key as identity claims, and the appraisal policy can require these claims to match trusted reference values. 
}
\changeB{
The \verifier Host establishes this identity claim independently of the \component and selects the policy and reference values without relying on component-emitted claims. A policy can therefore pin a \component hash or authorize a signer, preventing unauthorized replacement and cross-platform impersonation. When the policy pins a specific hash, it also prevents rollback to another version. A signer-only policy instead delegates version authorization to that signer, so component-signature expiry and revocation must be enforced.
\designname cannot detect an authorized but semantically faulty \component, or false endorsements issued by a compromised trusted implementer. Such failures affect attestation systems generally. An authorized component remains in the result-correctness \gls{tcb} for its platform, and its claims are attributable to the implementer that certified it.
}

\changeB{\textbf{Containment.}}
\changeB{
Regardless of whether the \component's output is authorized, \designname isolates the \component inside the WebAssembly sandbox and exposes only constrained host interfaces. 
\designname applies computational limits, per-component filesystem isolation, and an execution policy controlling network access.
Under the WebAssembly isolation assumption in Section~\ref{sec:adv-model}, a malicious or vulnerable \component cannot access the result-signing key, compromise the \verifier Host, interfere with other \components, or cause its output to be attributed to another \component.
The \components are therefore outside the host-security \gls{tcb}, although an authorized \component remains inside the result-correctness \gls{tcb} only for the platform whose evidence it interprets. Consequently, adding a new \component does not enlarge the \verifier Host's \gls{tcb} or the result-correctness \gls{tcb} of existing platforms.
}

\changeB{
Fuel metering bounds computation per invocation, while removing network access from unsigned or otherwise untrusted \components limits network abuse. These mechanisms do not prevent an adversary from submitting an unbounded number of requests; repeated-request denial of service requires admission control or rate limiting at the surrounding service.
}

\changeB{\textbf{\gls{tcb} and residual risk.}}
\changeB{
Both Trustee~\cite{confidentialcontainers2025trustee} and Veraison~\cite{ccc2023veraison} include platform-specific parsing and cryptographic logic in the result-correctness trust boundary. Trustee executes this logic directly within the Verifier process, while Veraison executes plugins as separate native processes communicating with the Verifier over RPC. This provides process-level isolation, but not the stronger sandboxing used by \designname, where WebAssembly confines \components, network access is controlled by execution policy, filesystem access is isolated per \component, and execution is bounded by fuel.
}

\changeB{
Currently, Trustee ships 11 native verifier backends. 
This understates its true trust boundary because these backends link substantial memory-unsafe native libraries that run with full host privilege, including OpenSSL (whose parsing code has historically yielded remote-code-execution vulnerabilities~\cite[CVE-2012-2110]{openssl-vulnerabilities}), and Intel's DCAP Quote Verification Library.  
Veraison~\cite{ccc2023veraison} independently maintains an overlapping set of verifier backends (both ship separate \gls{sev-snp} and Arm \gls{cca} verifiers, for instance), creating duplication effort that \designname avoids by allowing the logic to be authored once reused.
This per-platform native code and its dependencies grow without bound as new \glspl{tee} are supported, and a vulnerability in any one backend, or in any of these shared native libraries, can compromise the verification of all platforms.
\designname removes this cross-platform coupling. A \component is in the result-correctness \gls{tcb} only for the platform whose evidence it verifies; \components for unrelated platforms are outside that trust boundary.
}

\begin{figure}[t]
   \centering
   \includegraphics[width=\linewidth]{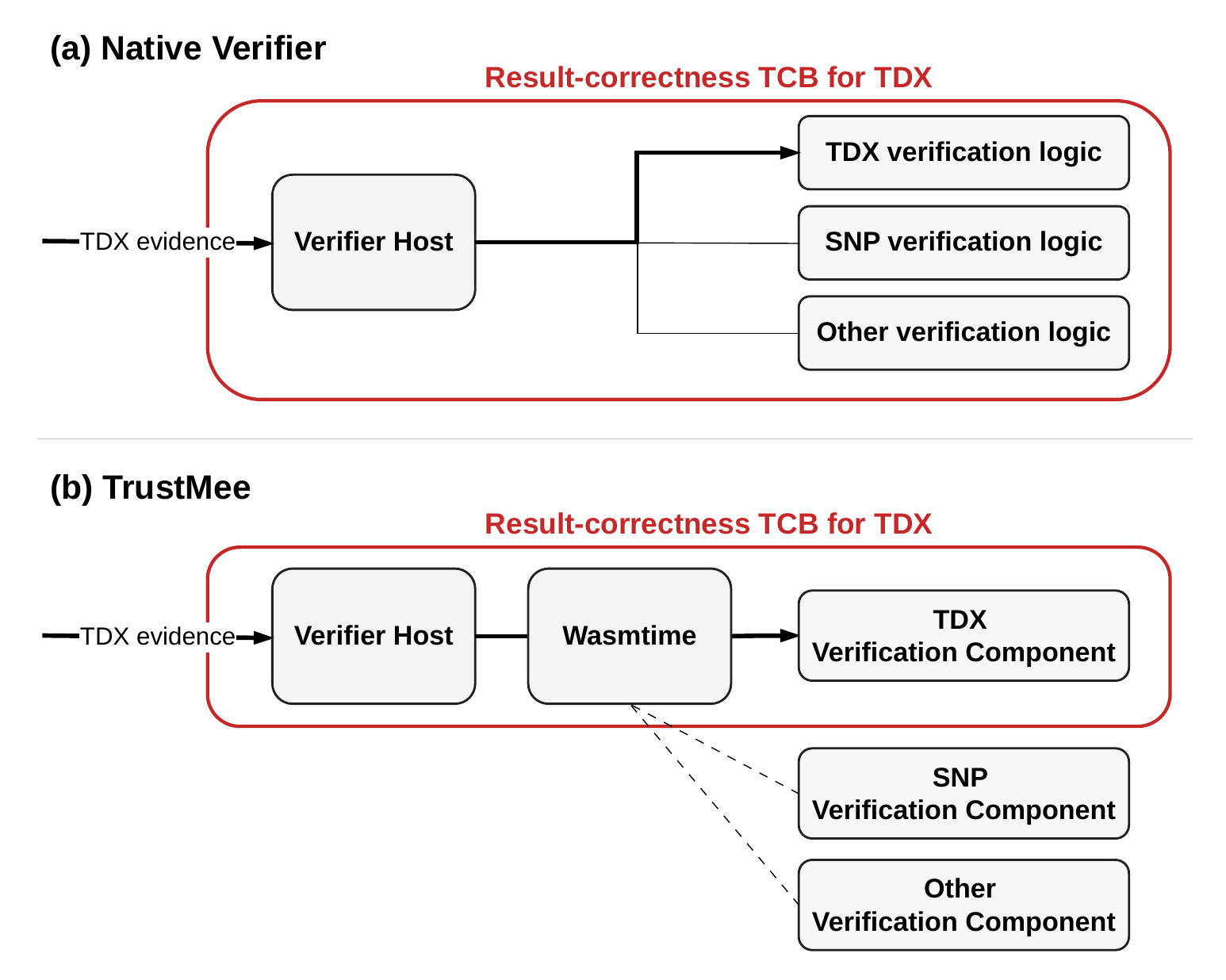}
   \caption{\changeB{Result-correctness \gls{tcb} comparison between native \verifiers and \designname. \designname adds Wasmtime as a fixed \gls{tcb} cost while keeping \components for unrelated platforms outside the result-correctness \gls{tcb} required to verify the selected platform. \gls{tdx} verification is shown as an example.}}
   \label{fig:trustmee-tcb}
 \end{figure}

\changeB{
\designname instead adds a single WebAssembly runtime to the host-security \gls{tcb}. We link Wasmtime~\cite{bytecodealliance2022wasmtime}, using only the Cranelift compiler backend targeting \texttt{x86\_64}. 
Although the runtime adds a fixed cost to the \gls{tcb}, this cost is identical whether \designname supports one Attester platform or fifty because the runtime is shared across every platform.
Crucially, from the host's perspective, the marginal \gls{tcb} cost of supporting a new platform is zero.
This is illustrated in~\Cref{fig:trustmee-tcb}.
}

\changeB{
As assumed in Section~\ref{sec:adv-model}, the Wasmtime runtime and Cranelift compiler are trusted. Their residual sandbox-escape risk therefore remains part of the Host \gls{tcb}.
}

\changeB{
\designname uses Wasmtime's default \texttt{x86\_64} Cranelift. Since Cranelift development began, Cranelift's public security documentation lists two sandbox-escape-capable miscompilation CVEs on x86/x86-64: CVE-2021-32629 and CVE-2023-26489~\cite{cranelift2026security,wasmtime2021cve32629,wasmtime2023cve26489}. Despite these historical vulnerabilities, Cranelift's correctness is increasingly machine-checked. Its instruction-lowering rules have been formally verified against SMT-based semantics~\cite{vanhattum2024lightweight}, and its register allocation is validated by a symbolic checker under continuous fuzzing~\cite{cranelift2026security}. 
These measures support, but do not eliminate, the assumption that malicious \components cannot escape the sandbox.
}

Based on this analysis, \designname satisfies our security requirement (SR, Section~\ref{sec:requirements}) under the stated adversary model (Section~\ref{sec:adv-model}).

\subsection{Performance Evaluation}

To evaluate the performance overhead of \designname in the Trustee Attestation Service, we compare the latency of the baseline Trustee Attestation Service, implemented fully in native code, with that of \designname, an Attestation Service that uses WebAssembly-based \components for evidence validation.
\changeB{We report performance comparisons only for the hardware-platform \components, for which Trustee provides equivalent native implementations that serve as meaningful baselines.}
For Intel \gls{tdx} and Intel \gls{sgx}, we further compare \designname against Trustee with native \verifier Drivers using the pure-Rust \texttt{dcap-qvl} instead of Intel DCAP QVL, as described in Section~\ref{sec:tdx-verifier-design} and~\ref{sec:sgx-verifier-design}.

\emph{Latency} captures the time required to handle an attestation request and carry out its key steps. We focus on \emph{end-to-end attestation verification latency} as the total time from when the HTTP request is received by the Trustee Attestation Service until the signed attestation result is returned to the client.

In our experiments related to \designname, we distinguish between cold and warm starts. By cold start, we mean that no compiled \component is cached in memory, although Wasmtime's file-based compilation cache remains enabled. We define a memory-warm start as the case where the compiled \component is cached in memory, but has not yet been instantiated.

We report mean and standard deviation of end-to-end attestation verification latency and verifier-specific computation time for all experiments over 50 runs per setting unless stated otherwise.

\subsubsection{Experiment Setup}
All experiments are executed inside a Docker container on the host OS. The container OS was Debian GNU/Linux 13 (trixie) running on an Ubuntu 24.04.3 LTS (Noble Numbat) host. The machine was x86\_64 with an Intel Xeon Silver 4510T, 24 logical CPUs (12 cores, 2 threads/core, up to 3.7 GHz), and 251~GiB RAM.

The WebAssembly runtime used is Wasmtime v41.0.4. We use Python scripts to send HTTP requests to the Attestation Service.

 \begin{figure}[t]
   \centering
   \includegraphics[width=\linewidth]{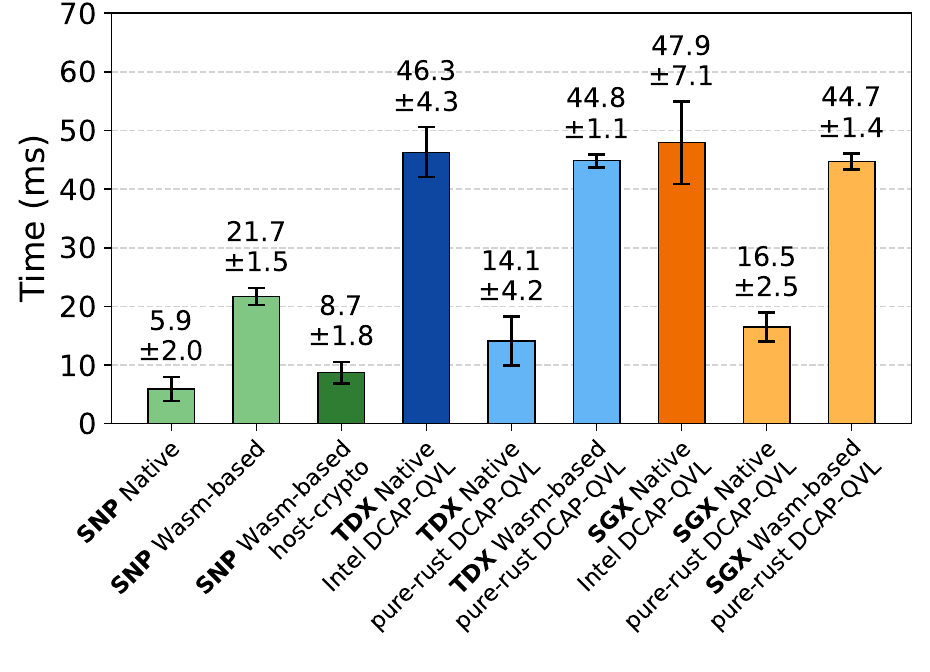}
   \caption{End-to-end attestation verification latency without network overhead (AMD \gls{sev-snp}, Intel \gls{tdx}, and Intel \gls{sgx}). Error bars are 1~s.d.}
   \label{fig:latency-with-cache}
 \end{figure}

\begin{figure}[t]
  \centering
  \includegraphics[width=\linewidth]{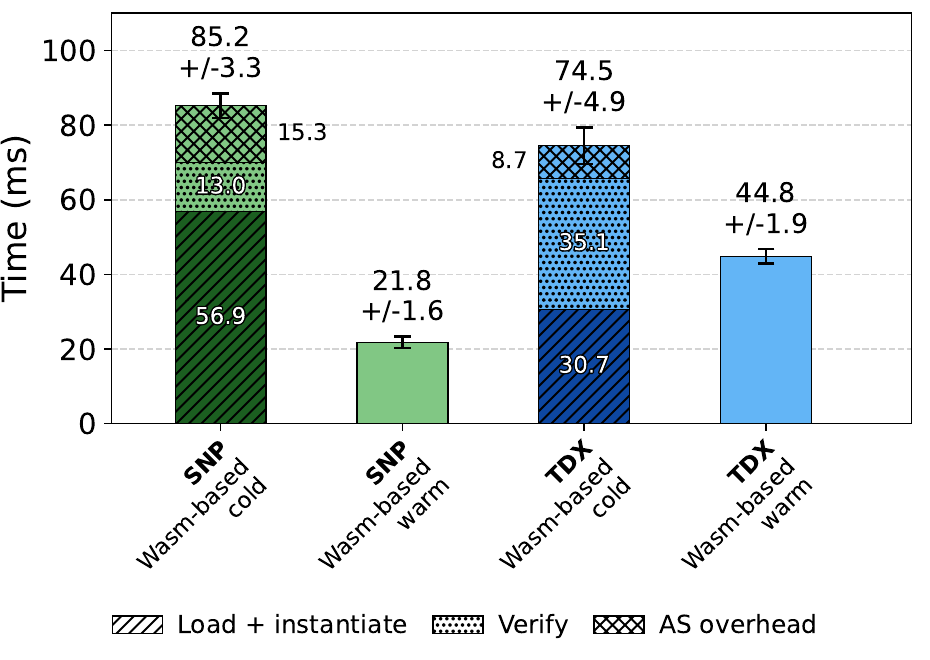}
  \caption{Effect of \component caching in \designname. AMD \gls{sev-snp} and Intel \gls{tdx} \component performance is measured in \designname without any in-memory caching (memory-cold start), and with in-memory caching of the compiled code (memory-warm start). Wasmtime's persistent compilation cache remains enabled in both cases. Error bars are 1~s.d.}
  \label{fig:wasm-cold-hot}
\end{figure}

\subsubsection{End-to-End Attestation Verification Latency}

\changeB{
We measure latency from receipt of the HTTP request to return of the signed EAR. To isolate \designname's computational overhead, both collateral and the \component are locally available. 
A warm-up request populates the component cache, \gls{sev-snp} certificates are embedded in the evidence, collaterals for \gls{tdx} and \gls{sgx} are stapled as a CMW endorsement, and measured network time is subtracted from the native \gls{tdx} and \gls{sgx} paths since they lack a mechanism for cached collaterals.
}

\changeB{
Figure~\ref{fig:latency-with-cache} reports warm-start results. The \gls{sev-snp} WebAssembly \component has $3.67\times$ the native end-to-end latency, consistent with previously reported WebAssembly cryptographic slowdowns~\cite{spies2025exploring}. Most of this difference comes from compiling native-oriented cryptographic code to WebAssembly. The \gls{tdx} and \gls{sgx} components are respectively $3.17\times$ and $2.7\times$ slower than native drivers using the same pure-Rust \texttt{dcap-qvl}, but are 3.3\% and 6.7\% faster than Trustee’s original Intel-QVL paths. Thus, \texttt{dcap-qvl} substantially improves native verification, while compiling that path to WebAssembly introduces the remaining overhead; \gls{tdx} and \gls{sgx} behave similarly because their evidence and validation paths are structurally related.
}

\subsubsection{Impact of Host-Based Cryptography on Attestation Verification Latency}
\label{sec:host-cryptography}

\changeB{
To estimate the benefit of native cryptographic implementations exposed to the \component via host functions, an \gls{sev-snp} variant imports a host function for X.509 chain and signature verification. As shown in Figure~\ref{fig:latency-with-cache}, this reduces warm-start overhead from $3.67\times$ to $1.47\times$, reducing the  absolute increase to below 3 ms. This approximates the benefit expected from future Component-Model support in interfaces such as WASI-crypto~\cite{webassembly2020wasicrypto}.
}

\subsubsection{\changeB{Cold Starts and Network Access}}

\changeB{
A cold request may fetch the component from an OCI registry, compile it if absent from Wasmtime's persistent cache, and load and instantiate it. Figure~\ref{fig:wasm-cold-hot} separates verification, loading and instantiation, and other Attestation Service work for \gls{sev-snp} and \gls{tdx}; \gls{sgx} is omitted because its component size, verification path, and runtime behavior closely match \gls{tdx}. The persistent compilation cache remains enabled, so compilation is excluded from the bars and takes $513.5 \pm 11.3$ ms for \gls{sev-snp} and $272.1 \pm 4.5$ ms for  \gls{tdx}. In-memory caching removes retrieval, compilation, and loading from the per-request path, leaving only instantiation and verification on warm starts.
These \designname-specific costs occur only when a \component is first encountered, while collateral-fetch latency affects both \designname and native Trustee. Where network fetching is required, it is expected to dominate total latency.
}

\subsubsection{\change{Attestation Verification Request Size}}

\begin{table}[t]
\centering
\caption{\change{Attestation request size change relative to native request baselines.}}
\label{tab:request-size-increase}
\begin{tabular}{@{}lcrrl@{}}
\toprule
\textbf{TEE} & \textbf{Collateral} & \textbf{Native}
& \textbf{Comp. ID} & \textbf{Change} \\
\midrule
SNP & $\circ$
& 3,257 B
& 3,700 B
& +13.6\%, $\times$1.14 \\

SNP & $\bullet$
& 9,039 B
& 5,634 B
& -37.7\%, $\times$0.62 \\

TDX & $\circ$
& 9,030 B
& 7,206 B
& -20.2\%, $\times$0.80 \\

TDX & $\bullet$\rlap{\hspace{0.1em}\textsuperscript{\dag}}
& 37,895 B
& 28,918 B
& -23.7\%, $\times$0.76 \\

SGX & $\circ$
& 8,519 B
& 6,843 B
& -19.7\%, $\times$0.80 \\

SGX & $\bullet$\rlap{\hspace{0.1em}\textsuperscript{\dag}}
& 42,014 B
& 32,026 B
& -23.8\%, $\times$0.76 \\
\bottomrule
\end{tabular}

\vspace{0.4em}
\begin{minipage}{0.96\linewidth}
\footnotesize
$\bullet$ Collateral included; $\circ$ collateral not included.
\textsuperscript{\dag}The native TDX and SGX verifier paths do not currently support stapled collateral; their native with-collateral baselines and corresponding changes are estimates.
\end{minipage}
\end{table}

\changeB{
Table~\ref{tab:request-size-increase} compares \designname requests carrying a \component identifier with native Trustee requests.
In our implementation, \designname requests are smaller in five of the six evaluated configurations, with reductions of approximately 20--38\% in four cases. The difference is primarily due to request encoding. \designname carries binary values directly in CBOR, whereas Trustee's native JSON path uses Base64 encoding, including nested Base64 encoding for some binary values.
If Trustee's native path used an equivalent CBOR encoding, \designname would add only about 300~B for its identifier and metadata.
Stapling the complete \component instead of referencing a cached or OCI-hosted \component increases the request by approximately 3 to 4~MB.
}

\textbf{Meeting the performance requirement.}
\changeB{
Fetch and compilation costs are amortized through caching.
In the evaluated warm-start cases, \designname introduces no latency overhead for Trustee’s original \gls{tdx} and \gls{sgx} paths and is faster in both cases, while host-provided cryptography reduces the \gls{sev-snp} latency overhead relative to native Trustee to below 3~ms. 
\designname therefore satisfies the performance requirement (PR, Section~\ref{sec:requirements}).
}

\section{Discussion}

In this section, we reflect on the design, implementation, and evaluation of \designname.

\subsection{Limitations and Trust Provisioning}

\designname does not eliminate trust in platform-specific verification logic. An authorized but semantically faulty \component can still produce incorrect claims \emph{for its own platform}. The \verifier operator provisions trusted component-signer keys to the Host trust store, while the appraisal policy authorizes the component hash or signer identity included in the \gls{ear}. \designname does not define the \gls{pki} or key-management for deciding which signers to trust. The \designname Host relies on Wasmtime and Cranelift to provide isolation and to limit computation by the component itself. Traditional mechanisms such as rate limiting and IP blocking can be used to prevent denial-of-service attacks caused by excessive numbers of requests.

\subsection{\changeB{Attester platform} support}

\designname can support any Attester platform, including \gls{tee} platforms such as Arm \gls{cca}~\cite{arm2024cca}, whose verification library can be packaged as a WebAssembly \component. As the \designname \verifier is outside the Attester platform, the architecture of the Attester platform in question is immaterial to its use with \designname.

\subsection{\change{Verifier-agnosticism}}
A \designname plugin for Veraison~\cite{ccc2023veraison} would allow the same \components to be reused across Verifier frameworks, making self-verifying evidence agnostic to both the Attester platform and Verifier framework.

\subsection{\changeB{Layered and Composite Attestation}}

\changeB{
The \gls{rats} architecture describes layered attestation, where evidence represents nested attesting environments, and composite attestation, where evidence is produced by multiple Attesters~\cite{birkholz2023remote}. The \gls{rats} multi-verifier architecture generalizes both as Composite Attesters, allowing different Verifiers to appraise parts or layers before a Lead Verifier combines their results~\cite{ietf-rats-multi-verifier-00}. \designname realizes the same decomposition within one Verifier by composing \components that share an interface, with each handling only its evidence format or layer.
}

\changeB{
Our Kata policy \component (\Cref{sec:kata-verifier}) demonstrates this structure. It is statically composed with the \gls{tdx} or \gls{sev-snp} \component. The higher-layer \component verifies the Kata policy and its binding to the \gls{cvm}, while delegating hardware evidence verification to the lower layer. The same approach can combine independently developed \components for composite evidence without a monolithic implementation that understands every format.
}

\changeB{
This also separates higher-layer schemes from hardware verification. An application-level scheme such as S-FaaS~\cite{alder2019s} can add its own evidence and verification while delegating \gls{sgx} quote verification through the common interface to an independently maintained \gls{sgx} \component, allowing the layers to evolve independently.
}

\changeB{
Our current implementation performs this composition statically. A future extension could instead resolve \component dependencies dynamically. A higher-layer \component could request lower-layer verification through the common interface, while the Verifier Host selects an authorized implementation according to policy. A Relying Party could therefore require a particular \gls{tdx} verifier without modifying the higher-layer \component.
This would allow higher-layer verification logic to remain independent of the implementation used to verify the underlying platform.
}

\subsection{\changeB{Hybrid deployment with native \verifiers}}

\changeB{
Dynamic resolution of lower-layer verification also enables hybrid deployments in which the selected implementation need not always be a \component. A production deployment may retain native Verifier Drivers for a small set of widely deployed \glspl{tee}, while using \designname for less common, newly introduced, or higher-layer evidence formats. Our evaluation shows that \designname is already competitive with Trustee's shipped \gls{tdx} and \gls{sgx} paths, while host-provided cryptography reduces the \gls{sev-snp} overhead to below 3 ms.
In such a hybrid deployment, these native drivers form a fixed part of the Verifier’s host-security \gls{tcb}, while additional platforms supported through \designname do not enlarge it.
}

\subsection{\changeB{Confidential Verification Components}}

\changeB{
The \gls{rats} multi-verifier architecture motivates decomposition partly because a Verifier Owner may keep its appraisal policy for evidence internal and a Reference Value Provider may not wish to disclose its reference values or their lifecycle to a monolithic Verifier~\cite{ietf-rats-multi-verifier-00}. \designname provides a natural basis for such confidential appraisal through confidential \components.
}

\changeB{
A component owner could first attest a \designname Verifier running inside a \gls{cvm}, establish a secure channel to it, and provision an encrypted \component that is decrypted only inside the \gls{cvm}. The component can then perform verification and appraisal internally while exposing only the required result. The \gls{cvm} protects the component from the hosting infrastructure, while the WebAssembly sandbox isolates it from the Verifier Host. We leave confidential provisioning to future work.
}

\subsection{\change{Network access by unsigned Verification Components}}
\label{sec:zerotrust}

\changeB{
Only \components signed by a trusted authority may use WASI-HTTP to fetch endorsements or collateral; untrusted components must use stapled collateral. The runtime or Host could instead allowlist trusted endorser hosts, limiting a compromised component while preserving required functionality~\cite{deislabs2024wasiexperimentalhttp}, but this weakens platform-agnosticism. More flexible alternatives are server opt-in, for example through a special DNS record that includes the component's hash, or routing component network access through the Attester so abuse consumes the Attester’s resources and cannot amplify traffic.
}

\section{Related work}

\subsection{Vendor-specific attestation verification}
Hardware attestation has evolved from TPMs~\cite{specification2007architecture} to modern \glspl{tee}, but remains vendor-specific. Intel have used \gls{epid}~\cite{johnson2016intel} and later DCAP-based attestation for \gls{sgx}/\gls{tdx}~\cite{intel2025sgxdatacenter}, Arm provides \gls{psa} attestation model for Arm-based IoT~\cite{arm2024psatoken}, and AMD uses its own format~\cite{amd2022sevsnpabi}. Although secure, these mechanisms remain hardware-specific and create integration and compatibility challenges in heterogeneous deployments.
Major cloud providers also offer platform-specific attestation services, such as the ones for AWS Nitro Enclaves~\cite{aws2024nitroenclaves}, Google Cloud Confidential Computing~\cite{googlecloud_confidential_in_use}, and Microsoft Azure Confidential Computing~\cite{microsoft2023azurecc}. While robust, these solutions are tightly coupled to their respective platforms. In addition, Intel Trust Authority~\cite{intel2023ta} is a similar online attestation service that supports a specific set of platforms.  These services ease deployment of remote attestation functionality, but do not address its platform-tied nature.

\subsection{Cross-platform attestation verification}

Prior work has pursued hardware-agnostic attestation verification through standards and modular \verifier{}s. The IETF \gls{rats} architecture~\cite{birkholz2023remote} defines interoperable roles and protocols, but it does not remove the need for platform-specific verifier implementations. \gls{rats} \gls{eat}~\cite{rfc9711}, on the other hand, is a standardized format for claims about entities. Only some platforms, however, currently produce and sign their evidence in this format, and the tokens may contain, e.g., proprietary claims that the \verifier needs to be able to interpret.
\changeB{
Concretely, the main server \glspl{tee} do not emit \gls{eat} as their native evidence. Intel \gls{tdx} quotes and AMD \gls{sev-snp} reports remain platform-specific binary evidence structures. Existing \gls{eat}/CoRIM standardization efforts address adjacent layers, such as representing verifier-produced \gls{tdx} attestation results or \gls{sev-snp} reference values, and remain IETF Internet-Drafts~\cite{ietf2026tdxeat,ietf2026snpcorim}; they do not remove the need to parse and validate the native evidence formats. 
\changeB{CoRIM does not inherently increase policy-language expressiveness, as given the same authenticated inputs, a general-purpose engine such as Rego can implement equivalent appraisal logic; CoRIM instead standardizes how those inputs are represented, authenticated, and reconciled~\cite{ietf-rats-corim}.}
Moreover, standardization has not eliminated evidence diversity. Device attestation increasingly uses SPDM~\cite{dmtf2025spdm}, e.g., in NVIDIA GPU attestation~\cite{nvidia2026gpuattestation}, so a \verifier must support another protocol and evidence encoding in addition to CPU \gls{tee} evidence. 
Cloud deployments add further envelopes around the same underlying hardware evidence. For example, Azure confidential VMs expose guest attestation through a \gls{vtpm}-based design, producing TPM quotes, PCRs, and event logs on top of AMD \gls{sev-snp} or Intel \gls{tdx} hardware~\cite{microsoft2025azurecvmattestation}; correspondingly, Trustee treats bare \gls{tdx}, bare \gls{sev-snp}, Azure \gls{tdx}-over-\gls{vtpm}, and Azure \gls{sev-snp}-over-\gls{vtpm} as distinct supported platforms~\cite{confidentialcontainers2025trustee}. The verifier maintenance burden therefore scales not merely with the number of hardware platforms, but with the combinations of platforms, evidence formats, and deployment envelopes that it must support. \designname is designed to address this diversity by packaging platform-specific verification logic as reusable \components that can be fetched and executed on demand, and composed naturally for layered or composite Attester platforms.}

\changeB{Plugin-based \verifier{}s such as Trustee~\cite{confidentialcontainers2025trustee} and Veraison~\cite{ccc2023veraison} provide a unified framework or API, yet still require per-\gls{tee} plugins or }drivers that must be built, deployed, and maintained for every new platform. In these designs, \gls{tee}-specific parsing and cryptographic logic typically runs as native code on the \verifier host, increasing operational overhead and exposing the \verifier to vulnerabilities in platform-specific code and its native dependencies.

The fragmented nature of current attestation systems presents significant challenges in heterogeneous environments. Gu et al.~\cite{gu2022unified} propose UniTEE, a unified attestation framework that addresses interoperability between different \gls{tee} platforms through a microkernel-based design. While their work demonstrates the feasibility of unified attestation, it relies on native implementations of verification logic for each platform, requiring separate compilation and deployment processes.

Related efforts aim to unify \gls{tee} development and deployment, but attestation coverage remains limited in practice. Niemi et al.~\cite{niemi2022towards} survey open-source attempts and note the difficulty of achieving broad, stable multi-\gls{tee} support. Enarx~\cite{enarx2023technicalintro} uses WebAssembly for application portability, but its attestation support is still platform-specific and currently limited (e.g., \gls{sgx}/\gls{sev-snp}). Open Enclave~\cite{microsoft2019openenclave} offers a hardware-agnostic programming model, yet real-world support is constrained (e.g., \gls{sgx} and limited TrustZone/OP-TEE support~\cite{linaro2019optee}). Veracruz~\cite{ccc2021veracruz} also supported only a small set of \glspl{tee} before moving to emeritus status.

\subsection{Platform-agnostic attestation evidence}
Another direction is to standardize the attestation report itself. Ott et~al.~\cite{ott2023universal} propose a universal report format in which the \verifier validates signatures and compares measurements against CA-signed reference metadata, largely avoiding platform-specific logic on the \verifier side. However, this shifts key trust and operational burdens to the Attester ecosystem. The \verifier must still decide which CAs are authorized to sign reference values and track revocation and updates, and the Attester must maintain complete and up-to-date reference values for all deployed components, moving the maintenance cost from a single \verifier deployment to many Attesters.
\designname instead keeps native evidence formats and moves their platform-specific verification logic into sandboxed \components supplied with the evidence, shifting verifier-code maintenance to the parties that own those evidence formats.

\subsection{\changeB{Layered and higher-layer attestation}}
\changeB{
Verification above the hardware evidence already exists but is tied to individual stacks. 
Project Oak's Transparent Release binds a binary's measurement to a signed endorsement in a transparency log and checks it in Oak's own client-side \verifier library for Oak's own stack~\cite{projectoak2024transparentrelease}. 
Google Confidential Space verifies container-image signatures and emits signing keys as claims, enabling signer-based policy instead of re-pinning the image digest on every release, but with a proprietary \verifier~\cite{google2025confidentialspace}. 
Keylime appraises Linux IMA logs against allowlists through its dedicated \verifier~\cite{schear2016keylime}. 
Layered attestation is also well established. DICE chains identities across boot stages, with each layer measuring the next~\cite{tcg2020dice}, while Copland formally specifies how evidence from multiple layers is orchestrated and bundled~\cite{ramsdell2019copland}; however, both rely on layer-specific processing logic already implemented in the device or appraiser.
These systems therefore encode verification, layering, and composition as stack-specific paths. \designname instead ships it as sandboxed, composable \components, maintained by the party that owns the format, so layered and composite attestation is expressed structurally, matching the \gls{rats} multi-verifier model without changes to the verifier core~\cite{ietf-rats-multi-verifier-00}.
}

\subsection{\changeB{Executing untrusted verification code}}

\changeB{
Web browsers routinely execute untrusted code, and WebAssembly was originally designed for this setting, allowing code compiled from languages such as C and C++ to execute within the browser~\cite{haas2017bringing}.
\designname similarly uses WebAssembly to isolate attacker-supplied \components.
}
To prevent malicious \components from exhausting \designname's resources, their execution must be bounded. Smart-contract platforms face the same problem and use gas to bound computation while requiring users to pay for the resources they consume~\cite{ethereum}. Wasmtime provides a similar fuel mechanism~\cite[\texttt{Config::consume\_fuel}]{wasmtime-docs}, which \designname uses to bound each \component's execution. A practical deployment may additionally require admission or accounting controls to prevent repeated invocations.
\change{
Other isolation mechanisms could also be used. For example, the eBPF verifier uses static analysis to prevent access to resources outside the sandbox and enforces limits on program size, execution complexity, and functionality such as looping to reduce the risk of denial-of-service attacks~\cite{ebpf_what_is_ebpf}. However, satisfying its restrictions may require substantial restructuring of existing verifier code.
}

\section{Conclusion}

This research presents the concept of \textit{self-verifying remote attestation evidence} that includes platform-specific verification code as a WebAssembly component. \change{This approach transforms evidence verification into a platform-agnostic functionality that is implemented once for all platforms}. \designname prototypes this functionality as a platform-agnostic Trustee verification driver, and demonstrates this with self-verifying evidence for AMD \gls{sev-snp}, Intel \gls{tdx}, and Intel \gls{sgx}\changeB{, as well as for two software-layer schemes, Kata agent-policy appraisal and ReCFA control-flow attestation, that existing \verifiers cannot express without modification.}

\changeB{Because the \verifier host contains no platform-specific code, \designname does not enlarge the host-security \gls{tcb} as platforms are added.
Compared with Trustee's native drivers, warm-start latency matches the shipped \gls{tdx} and \gls{sgx} paths, and \gls{sev-snp} overhead falls below 3 ms when cryptography is provided by the host. The remaining overhead stems from WebAssembly's limited support for cryptographic operations, and should diminish as interfaces such as WASI-crypto mature.}

In conclusion, our work demonstrates the feasibility of platform-agnostic attestation verification despite distinct evidence formats and verification workflows across hardware platforms.
\changeB{
By bundling verification logic with the evidence, \designname gives the \verifier the right logic at the right time, and lets whoever owns an evidence format determine how it is verified.}

\ifdefined\anonymous
\else
\section*{Acknowledgments}

This work has received funding from the Smart Networks and Services Joint Undertaking (SNS JU) under the European Union's Horizon Europe research and innovation program under Grant Agreement No 101139067. Views and opinions expressed are, however, those of the author(s) only and do not necessarily reflect those of the European Union. Neither the European Union nor the granting authority can be held responsible for them.

\section*{Contributions}

Z.A.~developed the concept and initial proof of concept~\cite{ahmad}, W.X.~developed the Trustee- and component-model-based architecture, interfaces, and initial prototype for \designname~\cite{xie}, P.S.S.~developed the public prototype and carried out the experiments, P.S.S., L.J.G., and M.S.~wrote the paper, and L.J.G.~and J.K.~supervised the work.

\section{Open Science}

The artifacts in this paper have been published on GitHub and have been made available at \url{https://github.com/elasticproject-eu/trustmee-artifact/tree/ndss}.
They include the \designname \verifier as well as all evaluation scripts needed to produce the figures and data in this work.

\fi

\bibliographystyle{IEEEtran}
\bibliography{sources}
\end{document}